\documentclass[preprint,12pt]{elsarticle}

\usepackage{amssymb}
\usepackage{amsmath}
\usepackage{courier}
\usepackage{subcaption}

\journal{Computer Physics Communications}

\begin{document}

\begin{frontmatter}


 \title{PULSEDYN - A dynamical simulation tool for studying strongly nonlinear chains}
%  \author{Rahul Kashyap and Surajit Sen\corref{cor1}\fnref{author2}}
% %% \ead{email address}
% %% \ead[url]{home page}
% %% \fntext[label2]{}

%  \address{Department of Physics, State University of New York at Buffalo, Amherst, NY 14260, USA\fnref{author1}}

%%\title{PULSEDYN - A dynamical simulation tool for studying strongly nonlinear chains}

 \author[label1]{Rahul Kashyap\corref{cor1}}
 \author[label1,label2]{Surajit Sen}
 \address[label1]{Department of Physics, State University of New York at Buffalo, Amherst, NY 14260-1500, USA}
 \address[label2]{Department of Physics, Brock University, St. Catharines, Ontario, Canada L2S 3A1}
 \cortext[cor1]{Corresponding author. \\ E-mail address: rahulkas@buffalo.edu}
% \author{R. Kashyap and S. Sen}

% \address{Department of Physics, State University of New York at Buffalo, Buffalo, New York 14260-1500, USA}

\begin{abstract}
We introduce PULSEDYN, a particle dynamics program in $C++$, to solve many-body nonlinear systems in one dimension. PULSEDYN is designed to make computing accessible to non-specialists in the field of nonlinear dynamics of many-body systems and to ensure transparency and easy benchmarking of numerical results for an integrable model (Toda chain) and three non-integrable models (Fermi-Pasta-Ulam-Tsingou, Morse and Lennard-Jones). To achieve the latter, we have made our code open source and free to distribute. We examine (i) soliton propagation and two-soliton collision in the Toda system, (ii) the recurrence phenomenon in the Fermi-Pasta-Ulam-Tsingou system and the decay of a single localized nonlinear excitation in the same system through quasi-equilibrium to an equipartitioned state, and SW propagation in chains with (iii) Morse and (iv) Lennard-Jones potentials. We recover well known results from theory and other numerical results in the literature. We have obtained these results by setting up a parameter file interface which allows the code to be used as a black box. Therefore, we anticipate that the code would prove useful to students and non-specialists. At the same time, PULSEDYN provides scientifically accurate simulations thus making the study of rich dynamical processes broadly accessible. 

\end{abstract}


\end{frontmatter}

%\linenumbers


%Description of your software in maximum 5 pages for first Original Software Publication â- see suggested %format; 

Program Summary \\

\noindent Title: PULSEDYN \\
Licensing provisions: GNU General Public License 3 (GPL) \\
Programming Language: C++ \\
Supplementary Material: user manual, documentation \\
Operating Systems: Windows, Linux \\

Memory required: Dependent on system size. For 5000 particle chain around 2MB is required. \\

Running time: Dependent on simulation parameters. For a 100 particle chain with polynomial potential run time was 20 min to run over $2\times 10^6$ iteration while writing data every 1000 iterations.\\

Nature of the problem: We solve a Newtonian particle dynamics problem in one dimension for a variety of commonly used nonlinear manybody systems. The goal is that the user should be able to run scientifically accurate simulations with minimal effort. Benchmarking should also be easy to perform by different users running simulations across different platforms. \\

Solution method: The code uses a parameter file interface with a set of commands for a range of pre-built functionalities. By providing the code and the needed parameter file to the user we hope to make scientifically accurate simulations easier to perform. We provide the Gear $5^{th}$ order predictor-corrector algorithm as well as the velocity-Verlet algorithm and the following potentials - the Toda potential, Morse Potential, Lennard-Jones Potential and a polynomial potential up to $4^{th}$ order. To make benchmarking consistent we release the software as open source and free to distribute. \\

\section{Introduction}
\label{introduction}

The importance of computational power and techniques to analyze many-body nonlinear systems (MBNS) simply cannot be overstated. Major discoveries in many-body physics such as the famous Fermi-Pasta-Ulam-Tsingou (FPUT) paradox \cite{fermiLANLR1955,fermiChicago1965,todaJPSP1967_1} have been explored via numerical means. Sometimes, numerical techniques have been used in conjunction with analytical descriptions to provide evidence of phenomena \cite{todaJPSP1967_2,flachPhysRep1998,flachPhysRep2008,bermanChaos2005,campbellChaos2005,fordPhysRep1992,saitoJPSP1967}. However, a great number of times, numerical solutions are our only tool to explore the physics of a system when analytical techniques become unwieldy \cite{reigadaPRE2001,reigadaPhysicaA2002,sieversPRL1988,satoNature2004,bennetinJSP2011,takatoEPL2012,avalosPRE2011,senPhysRev2009,senPhysRep2008}.

There is significant effort and interest in developing a particle dynamics (PD) code to accurately solve coupled dynamical equations in MBNS over long time scales while retaining strict control over error propagation. PULSEDYN as a tool provides a solution to this issue and can be used readily to carry out detailed investigations of important MBNS such as the Toda chain system \cite{todaJPSP1967_2}, polynomial potentials, Morse potential and the Lennard-Jones potential \cite{flytzanisJPA1989}. The differential equation solvers currently available in the code are the Gear $5^{th}$ predictor-corrector method \cite{gear1966} and the velocity-Verlet algorithm \cite{swopeJCP1982} which is a commonly used modification of the St\"ormer-Verlet algorithm \cite{verletPR1967,stormer1921}. Interested readers can find comparisons of these algorithms in standard textbooks such as \cite{allen1987}.

PULSEDYN currently is available for Windows (testing performed on Windows 10) and Linux (testing performed on Ubuntu 14.04 LTS) and we plan to add MacOS support in the future. In addition to containing the potentials and solvers listed above, it also contains features such as external forcing and dissipation \cite{satoNature2004}. It would allow one to vary a wide variety of parameters and set up highly accurate simulations with minimal effort. The code is also open source. Hence users with experience in $C++$ will be able to modify the same for specific purposes across different platforms. Further, we believe it could serve as a tool to benchmark other codes designed for the same purpose. 

In Section \ref{scope} we discuss the class of problems that the program is designed to solve. We then describe the details of the programming and how to  implement the same in Sections \ref{architecture} and \ref{functionality}. We show results obtained from the code in Section \ref{results} for test cases of the potentials built into the system. We test the four potentials provided in the program and show that the numerical results match the results in the literature. First, we show results for the Toda chain system which is integrable and compare the well known Toda soliton solution from simulations to theoretical predictions \cite{todaJPSP1967_1,todaJPSP1967_2,toda1981}. Here by soliton we mean a non-dispersive energy bundle that {\it does not} interact with another identical energy bundle except for a trivial phase lag that is induced by a slowdown during the collision, which is a property of integrable systems. Next, we consider non-integrable systems starting with the FPUT system and show the recurrence phenomenon, localized excitations and SW collisions in the lattice \cite{fermiLANLR1955,fermiChicago1965,kashyapIJMPB2017,zhaoPRL2005}. We then show SW propagation through the Morse and Lennard-Jones lattices and demonstrate that the simulations done using PULSEDYN are in agreement with established results from the literature \cite{flytzanisJPA1989}. Typically, SWs interact when they collide \cite{zhaoPRL2005,avalosPRE2009,kmPramana2005}. Finally, in Section. \ref{journeyToEqb}, we turn our attention to a problem of historical importance - namely equipartition of energy in the FPUT system \cite{fermiLANLR1955,fermiChicago1965}. We show that the results we have obtained using PULSEDYN demonstrate conclusively that the purely nonlinear $\beta$-FPUT system goes to equilibrium at late times and energy is equipartitioned in the system \cite{liviPRA1985,onoratoPNAS2015,benettinJSP2013,fergusonJCP1982,henonPRB1974}

\section{Problems and Background}
\label{scope}

MBNS are set up by defining a Hamiltonian and by then writing down the corresponding force equations. This gives rise to a set of coupled differential equations which are solved numerically using a well chosen integration algorithm depending on the problem at hand \cite{allen1987,press1988}. We have included the following potential functions into PULSEDYN, \\ 

\indent a)	Toda Potential \cite{todaJPSP1967_1,todaJPSP1967_2}, 
    \begin{eqnarray}
    V_{i, i+1} & = & \frac{k_1}{k_2} e^{-k_2(x_{i+1} - x_{i})} + k_1(x_{i+1} - x_{i}) - \frac{k_1}{k_2}, 
    \label{todaPotential}
    \end{eqnarray}
\indent b)	FPUT Potential ($\alpha$ + $\beta$ model) \cite{fermiLANLR1955,fermiChicago1965}, where $\alpha$, $\beta$ were used in the original and many subsequent works to denote the prefactors of the cubic and quartic terms in the potential, respectively,
    \begin{eqnarray}
    V_{i, i+1} & = & k_1(x_{i+1} - x_{i})^2 + k_2(x_{i+1} - x_{i})^3 + k_3(x_{i+1} - x_{i})^4,
    \label{fputPotential}
    \end{eqnarray}
\indent c)	Morse Potential \cite{flytzanisJPA1989,morsePR1929},
    \begin{eqnarray}
    V_{i, i+1} & = & k_1 (e^{-k_2(x_{i+1} - x_i)} - 1)^2,
    \label{morsePotential}
    \end{eqnarray}
\indent d)	and the Lennard-Jones Potential \cite{flytzanisJPA1989,lennardjonesPRSL1924},
    \begin{eqnarray}
    V_{i, i+1} & = & k_2 \Bigg[\Bigg(\frac{k_1}{k_1 + x_{i+1} - x_i}\Bigg)^{12} - 2\Bigg(\frac{k_1}{k_1 + x_{i+1} - x_i}\Bigg)^6 + 1\Bigg].
    \end{eqnarray}
In the potentials listed above, $k_1, k_2$ and $k_3$ ($k_3$ is only in Eq. \ref{fputPotential}) set the parameters of the potential. In the case of the Lennard-Jones potential, $k_1$ sets the bond length in the chain explicitly so care must be taken with initial conditions when setting the energy scales of the system. In the other potentials, bond length is not explicitly specified and the scales can be set arbitrarily. Additionally, in the case of the FPUT potential, the cubic term can cause instabilities if set incorrectly. So care must be taken to provide appropriate values of $k_2$ when using the FPUT potential. In our studies, we have set only positive $k_1, k_2$ and $k_3$. However, negative values of the parameters are not necessarily unphysical. Therefore, the code does not prohibit the user from setting negative parameter values. However, care must be taken when setting the values of the parameters and initial conditions so that the simulations remain stable.

Some discussion about numerical instability is warranted here. If two particles come too close to each other the magnitude of force can increase drastically. This can cause numerical instability due to floating point calculation errors if the parameters are not chosen carefully. For instance, if the energies are set too high, the time scales of interest in the dynamical studies may become too small to be resolved by the time step set in the parameter file. This would cause the simulation to become unstable. Another source of error could be the bond length being set too low for a given amount of energy. In PULSEDYN, If the bond lengths are set to be too small, particles can try to cross each other. This can lead to instability since particle crossover is not physical and hence not realizable without incurring errors in this code. For numerical stability, $|x_{i+1} - x_i| < |k_1|$ for the Lennard-Jones potential at all times during the simulation.

In addition, some problems also involve an external driving force and can include dissipation. The code solves the following force problem

\begin{eqnarray}
\frac{m_id^2 x_i}{dt^2} & = & -V'(x_i - x_{i-1}) + V'(x_{i+1} - x_i) + F_{dissipation} + F_{external}.
\label{forceEqn}
\end{eqnarray}

Here, $m_i$ and $x_i$ are the mass and displacement of the \textit{i}th particle. $V'(x_{i+1} - x_i)$ is the derivative $dV(r)/dr_i$ where $r _i = x_{i+1} - x_{i}$.  

The user can add external forcing and dissipation in the form of sinusoidal forces and velocity dependent damping terms respectively. These additional terms may be added to individual particles as well as to the entire system. We have also provided a method of external forcing called \textit{chirping}. Under the right circumstances, it can help precipitate intrinsic localized modes or localized nonlinear excitations \cite{sieversPRL1988,satoNature2004}.

\section{Code Architecture}

\label{architecture}

The code is written in an object oriented style to allow for good organization and legibility. The code contains 11 source file and 11 header files and the list is provided in the documentation pdf accompanying the software. It is distributed as a Code::Blocks project fully compiled and accompanied by an executable. 

Any user with a basic knowledge of $C++$ but lacking time to develop an entire program from scratch, can modify and add more features to these source files and customize it. The software is generous with comments in the source files. Every effort has been made to ensure the implementation is clear and self-explanatory. We request any user to give attribution to this work when reporting results using PULSEDYN or any edited version of the same.

The implementation also emphasizes encapsulation. For, instance, a user need only add to modify the boundary condition files and update the input file interface in order to impose custom boundaries on the system. The rest of the files need not be modified. More experienced users of $C++$ may of course be willing to modify the software extensively to customize it to their needs including optimizations for speed, additional models, algorithms etc. A detailed documentation of all the features is also provided with the code. The organization of the code is described next.

\subsection{Particle class}

The \textit{Particle} class, as the name suggests, defines a particle in a chain. A \textit{Particle} object carries the following information about each particle in the chain, \\
\noindent a) Position, \\
b) Velocity, \\
c) Acceleration, \\
d) Mass, \\
e) Kinetic energy and \\
f) Boundary information (useful only for boundary particles). \\

The following functions are included in the Particle class \\
\noindent a) Getter and setter functions for each member, \\
b) Kinetic energy calculator \\

\begin{figure}
\centering
\includegraphics[scale=0.24]{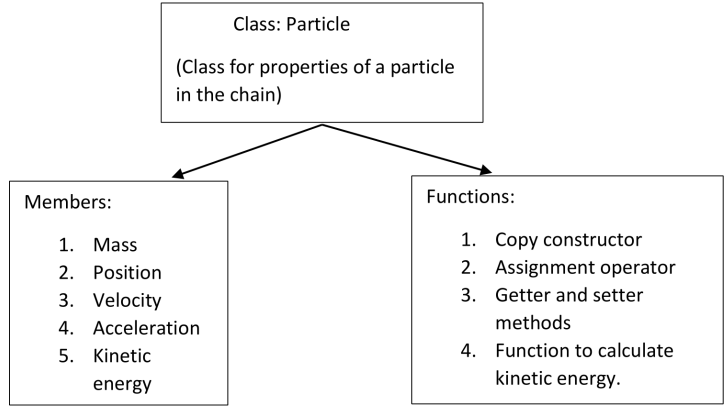}
\caption{Shown here is the \textit{Particle} class layout in the software.}
\label{parClass}
\end{figure}

For a chain of particles, a $C++$ vector of such objects is created and the dynamical quantities are updated in each iteration of the algorithm. Further details are given in Section \ref{simulation}. The organization of the class is shown in Fig. \ref{parClass}. 

\subsection{Potential classes}

The \textit{System} class is an umbrella class that contains all the potentials which are written as subclasses of the \textit{System} class. The information about the parameters in the potential are stored by an object of any of these subclasses. The functions for force and potential energy corresponding to each of these models are written as member functions of their own classes. There are 4 classes for each of the four potentials provided in the software. The classes are \textit{todaPotential, fpuPotential, morsePotential} and \textit{lennardJonesPotential}. An object of any of the \textit{Potential} classes stores the following members, \\
\noindent a) Parameters of the potential and\\
b) Potential energy. \\

Each of the potential classes contains the following functions, 
\noindent a) Getter and setter functions for each member, \\
b) Acceleration function  and\\
c) Potential energy function. \\

\begin{figure}
\centering
\includegraphics[scale=0.4]{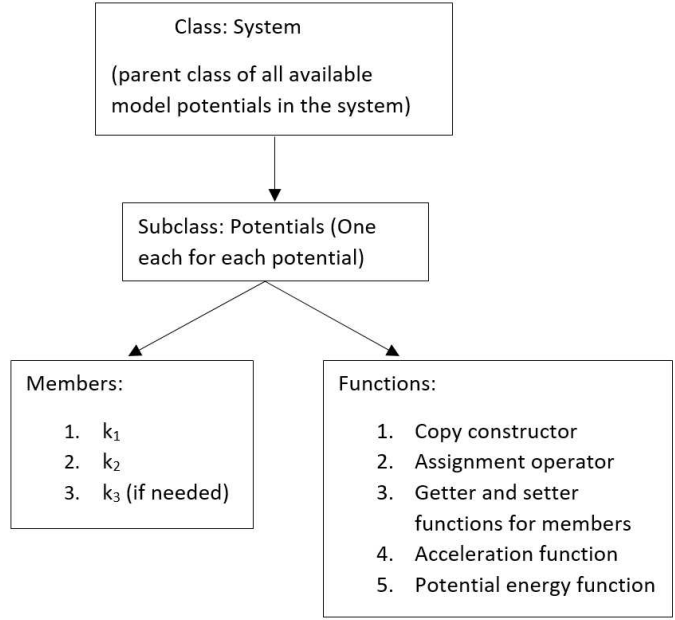}
\caption{Shown here is the layout for each of the \textit{Potential} classes in PULSEDYN.}
\label{sysClass}
\end{figure}

The acceleration function calculates the right hand side of Eq. \ref{forceEqn} excluding the driving and dissipation forces i.e. the force due to the potential. When required, the function is able to call the boundary conditions functions detailed below to calculate the accelerations for the edge particles. The potential energy calculates potential energy for each spring in the system. Similar to the acceleration function, the boundary conditions are calculated via external functions for the boundary springs by calling them inside the potential energy function. The information about which boundary conditions are to be used is stored in the \textit{Particle} class detailed before. A notable aspect of the potential functions is that the names of the functions and variables defined within each of the potential class and the structure of the potential classes are identical. The only difference exists in the actual equations that calculate the acceleration and potential energy. The reason for this manner of implementation is explained in Section \ref{simulation}. The layout of the \textit{Potential} classes is shown in Fig. \ref{sysClass}.

\subsection{External force}

While the \textit{Potential} classes take care of the potential energy and the resulting conservative force part of the right hand side of Eq. \ref{forceEqn}, the \textit{Force} class provides the velocity dependent damping and external driving part of the right hand side of Eq. \ref{forceEqn}. In PULSEDYN, we have provided sinusoidal driving forces with the option of a frequency ramp to create chirping \cite{satoNature2004}. The \textit{Force} class members are \\
\noindent a) Start time, \\
b) Period of the periodic force, \\
c) End time of the force, \\
d) Frequency (calculated from period above), \\
e) Ramp (for chirping), \\
f) Amplitude of driving force and \\
g) Velocity dependent damping $\gamma$. \\

The frequency is calculated based on the time period of the force as $f_o = 1/T$ where $f_o$ is the frequency and $T$ is the time period. The dissipation is provided as a velocity dependent dissipation term given by $\gamma v$ where $v$ is the velocity and $\gamma$ is the damping coefficient. Chirping is a specific way of driving the system such that the frequency increases linearly as a function of time i.e., $f(t) = f_o + (ramp\times t)$. The member functions of the \textit{Force} class are \\
%RAHUL - THE QUANTITY $f_o$ IN THE LINE ABOVE HAS NOT BEEN DEFINED
\noindent a) Getter and setter functions for each member, \\
b) Force select function, \\
c) Sine function and \\
d) Cosine function. \\

\begin{figure}[h!]
\centering
\includegraphics[scale=0.35]{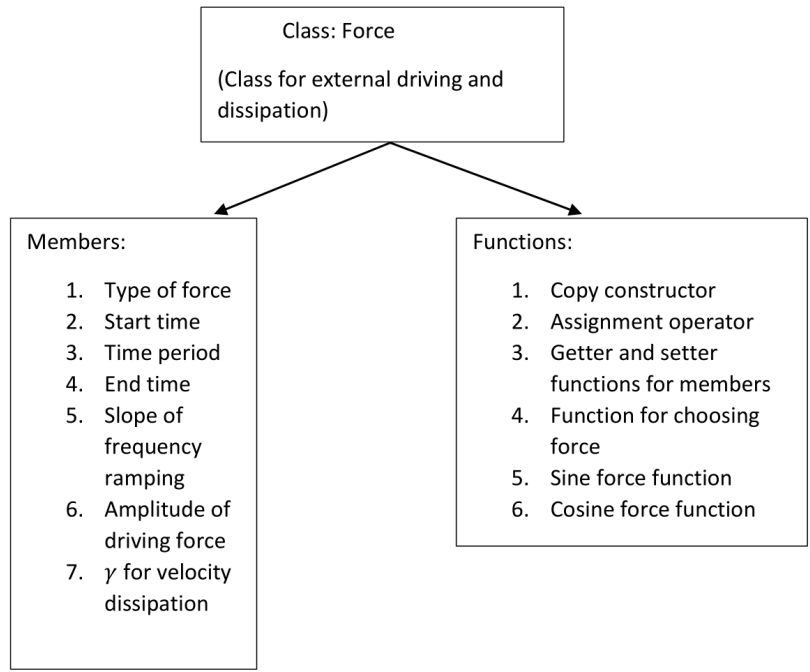}
\caption{Shown here is the layout for the \textit{Force} class in PULSEDYN.}
\label{forClass}
\end{figure}

It is important to note here that symplectic algorithms such as the velocity-Verlet only support equations of motion which are Hamiltonian \cite{press1988}. Adding external forces or dissipation would break the Hamiltonian nature of a system. Therefore, external driving and dissipation must be handled with care. More details about how this issue is handled are presented in Section \ref{simulation}. The organization of the \textit{Force} class is shown in Fig. \ref{forClass}.

\subsection{Simulation class}
\label{simulation}

\begin{figure}
\centering
\includegraphics[scale=0.4]{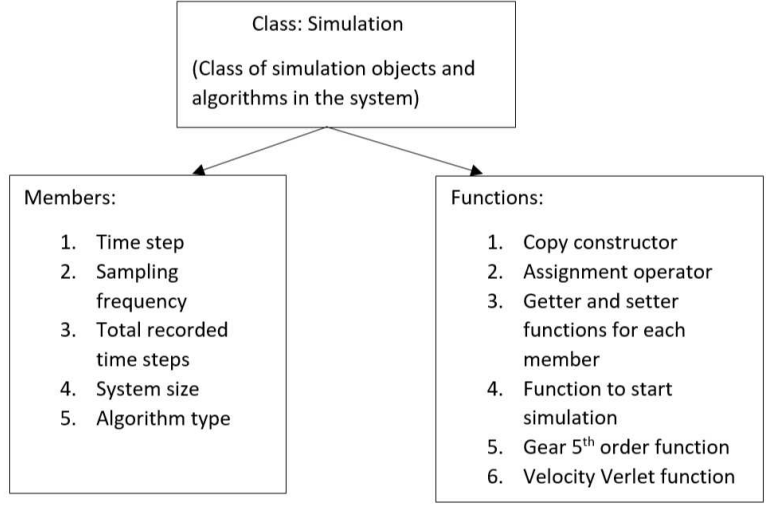}
\caption{Shown here is the layout for the \textit{Simulation} class in PULSEDYN.}
\label{simClass}
\end{figure}

The \textit{Simulation} Class contains the following members, \\
\noindent a) Time step for the integration algorithms, \\
b) Sampling interval, \\
c) Total number of snapshots in time recorded, \\
d) Size of the chain and\\
e) Integration algorithm name. \\ \\
The functions implemented in this class are as follows, \\
\noindent a) Getter and setter functions for each member, \\
b) Starter function for simulations, \\
c) Gear $5^{th}$ order predictor corrector algorithm and\\
d) Velocity-Verlet algorithm. \\ \\
The functions in the class are implemented as template functions. The functions take objects of generic type and the actual type is determined at runtime. Therefore, any object passed through the template function must have the acceleration and potential energy defined with the exact same name and have the same organization. We get three advantages from this approach. First, it enables code reusability. Second, the runtime efficiency of the code is better compared to the other options we can avail given the structure of this code. Third, it provides modularity to the code. If new potentials are added to the code, one need not modify the functions for the integration algorithms as long as the member functions and variables share the same names as the ones in the included potential classes. The organization of the \textit{Simulation} class is shown in Fig. \ref{simClass}.

The starter function in the simulation class checks which integration algorithm was chosen and directs the code to the chosen integration algorithm function. In each of the integration algorithms, the following routine is employed. First, the data before the simulation starts is written to file. Then the algorithm variables needed to execute the update steps are initialized. The function starts a time loop and the dynamical variables are updated in each iteration. During the update step, calculations of force and acceleration are performed to get the values of the position and velocity at the next iteration. The total force is calculated as a sum of the conservative and non-conservative parts of the right hand side (RHS) of Eq. \ref{forceEqn}. The conservative part is calculated by calling the acceleration functions inside \textit{Potential} class and the non-conservative parts are calculated by calling the functions in the \textit{Force} class. 

As remarked previously, the velocity-Verlet algorithm is unstable for non-conservative systems \cite{press1988}. Therefore, adding external forces to the system would cause the code to crash if velocity-Verlet is implemented without accounting for the external force. In PULSEDYN, \textit{if the user wishes to add external driving or dissipation to the system, the Gear $5^{th}$ order corrector-predictor method must be specified to integrate the equations of motion}. If the velocity-Verlet algorithm is called in the parameter file, the code ignores the non-conservative part of the equations of motion and solves the Hamiltonian part with the initial conditions and parameters as set in the parameter file. 

Inside the time loop, a counter is implemented which checks if the sampling interval is reached. When the writing condition is satisfied in the counter, the energies are calculated and the data is written to file. Once file writing is complete, the counter for file writing is reset. The details of file output are given in Section \ref{outputSection}.

\subsection{Boundary conditions}

Boundary conditions are implemented as stand-alone functions in the software. The functions for boundary conditions return the value of the spring extension or compression at the right end of the $N$th spring and the left end of the 1st spring. Three kinds of boundaries have been provided - open, fixed and periodic. The functions for boundary conditions are, \\
\noindent a) Left boundary function for acceleration, \\
b) Right boundary function for acceleration, \\
c) Left boundary function for potential energy and \\
d) Right boundary function for potential energy. \\

The functions take the vector containing the particle objects and evaluates the type of boundary chosen for both ends of the chain. Once the boundary type is evaluated, it sends back the spring extension or compression corresponding to the edge springs. In the case of the fixed end, the edge spring is fixed to a wall on one end and the boundary particle on the other. The functions then return the value of the spring extension or compression by assuming that the \textit{wall} displacement is zero at all time. If the boundary is open, the edge spring is connected only to the boundary particle i.e., the compression or extension in the boundary spring is always zero. It must be noted that imposing open boundaries or periodic boundaries can make the chain susceptible to drifting. For instance, a 2 particle simple harmonic oscillator with open ends would show drift as well as periodic oscillations if one of the particles is given an initial velocity perturbation in only direction. 

In the case of periodic boundaries, there is only one boundary spring and it connects particles 1 and $N$. Therefore, the magnitude of change in the length of the boundary spring is $|x_1 - x_N|$. In the case of the accelerations, the boundary conditions are calculated at both ends of the spring and used to calculate the forces on the boundary particles. However, calculating energies is trickier. To avoid counting the single periodic boundary spring twice, the boundary spring energy is calculated only at the left chain edge while the energy of the periodic boundary spring is calculated to be zero at the right end. This necessitates having two different boundary functions for acceleration and potential energy. It is also worth mentioning that in PULSEDYN, while the open and fixed boundaries can be set independently on the ends of the chain, periodic boundaries cannot. If one of the boundaries is set to be periodic the other boundary is also forced to be periodic. 

\subsection{Output class}
\label{outputSection}

To write data, we have implemented a static class \textit{Output} that can be called from anywhere in the code to write data to file. The following data is written to file, \\
\noindent a) {\tt position.dat} for displacement data, \\
b) {\tt velocity.dat} for velocity data, \\
c) {\tt acceleration.dat} for acceleration data, \\
d) {\tt ke.dat} for kinetic energy data, \\
e) {\tt pe.dat} for potential energy data, \\
f) {\tt restart.dat} records displacement, velocity and acceleration at the last time iteration, \\
g) {\tt mass.dat} for mass values of each particle and \\
h) {\tt totalEnergy.dat} for total energy at each recorded iteration. \\

The class contains the following members, \\
\noindent a) File names of data files and\\
b) File streams to data files. \\

The functions in the class are the WRITE function to the files. Each of the functions takes as arguments the value to be written to file and an end of line check. Once called, the functions open the associated files, write the value and either place a tab-space or go to the next line depending on the end of line argument. 

These functions are called from the UPDATE functions in the \textit{Simulation} class when the file writing counter determines that data writing condition is satisfied. The files are written for position, velocity, acceleration and kinetic energy as arrays of size $T \times N$, where $T$ is total recorded time steps and $N$ is the size of the chain. The mass of the particles is recorded as an $N \times 1$ list of numbers. The potential energy of the springs is recorded as a $T \times (N+1)$ array. As detailed previously, in the case of fixed boundaries at both ends, there are $N + 1$ springs. If one of the boundaries is open or if both are open, then there are $N$ and $N-1$ springs respectively. The code simply prints zeros on the side of the open boundary in the file. In the case of the periodic boundaries, there are $N$ springs. The spring connecting particle 1 and particle $N$ is considered the leftmost spring and the $N+1$th column in the file is given a value of 0 to avoid over counting. The code also prints out total energy of the system at each recorded step as an array of size $T \times 1$. 

\begin{figure}[h!]
\centering
\includegraphics[scale=0.4]{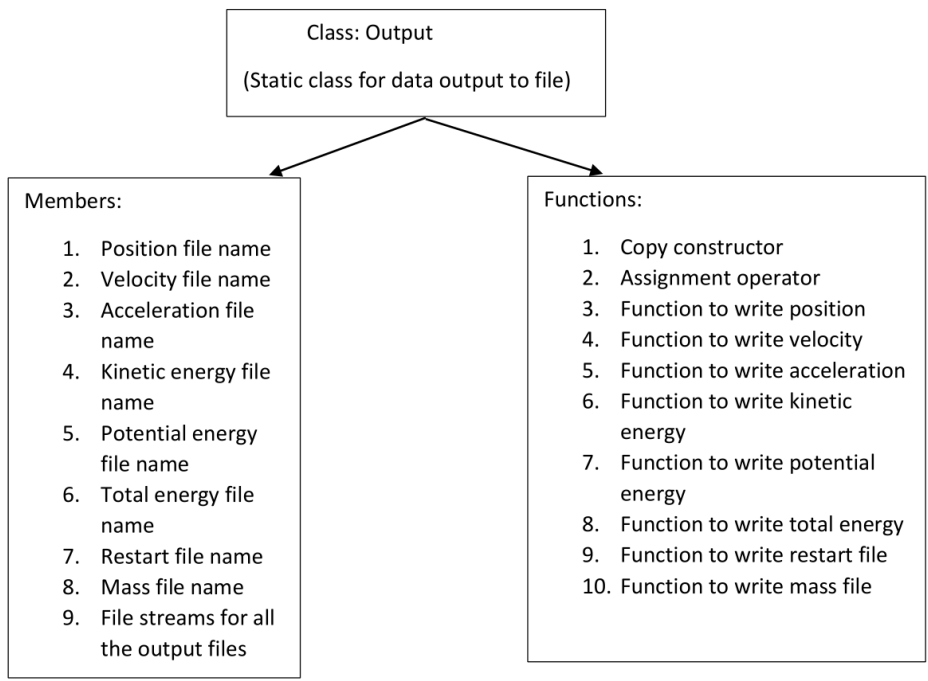}
\caption{Shown here is the layout for the \textit{Output} class in PULSEDYN.}
\label{outClass}
\end{figure}

The last file recorded is a restart file. It records the position, velocity and the acceleration as an $N \times 3$ array. The first column corresponds to position, the second to velocity and the third to acceleration. This is identical to the format required by the code to take initial conditions from file which was detailed earlier. The layout of the \textit{Output} class is shown in Fig. \ref{outClass}.

\subsection{Organization}
\label{organization}
 
\begin{figure}[h!]
\centering
\includegraphics[scale=0.4]{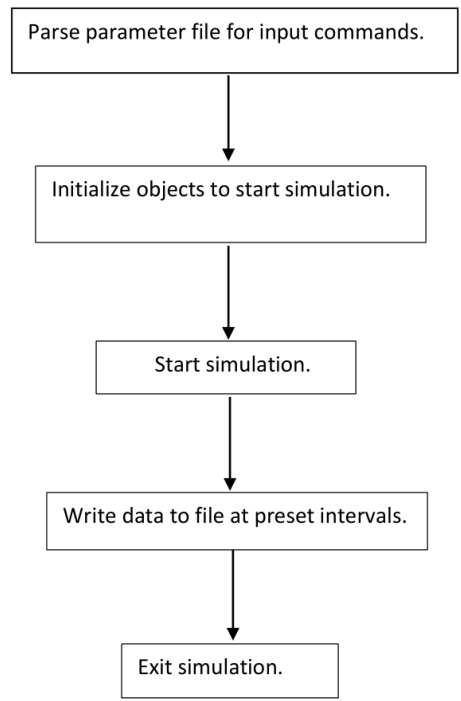}
\caption{A high level flowchart for the software is shown in this figure.}
\label{flowChart}
\end{figure}

 The code organization is straight-forward. First, the parameter file is parsed to extract values of the variables required for running a simulation. Any parameter not specified is simply set to its default value. The default values can be found in the user manual. Once the variables have been set, the objects for the classes in the software are initialized. After initialization, the model chosen is fed into the function that starts the simulation. Data is written to file periodically as explained in Section \ref{simulation}, till the simulation is over and the program exits. A flowchart illustrating the code organization is shown in Fig. \ref{flowChart}. 
 
\section{Software Functionalities}
\label{functionality}

The software can be run by setting up a few commands in the parameter file named \texttt{parameter.txt} and double clicking on an executable or by running the executable in a terminal window. A detailed explanation of how to specify these commands is given in a user manual provided with the software.

\begin{table}
\begin{center}
\begin{tabular}{ c c }
 \hline	
 \texttt{model:} & Specify potential type and parameters  \\ 
 \texttt{method:} & Choose integration algorithm \\  
 \texttt{systemsize:}& Set number of particles in the chain \\
 \texttt{timestep:} & Set the value of $dt$ used by the integration algorithm \\
 \texttt{recsteps:} & Set total number of recorded steps \\
 \texttt{printint:} & Set the time iterations after which data is recorded \\
 \texttt{init:} & Set initial conditions \\
 \texttt{boundary:} & Set boundary conditions \\
 \texttt{mass:} & Set masses of individual particles \\
 \texttt{force:} & Set external force on the system \\
 \texttt{dissipation:} & Set a velocity dependent dissipation \\
 \hline	   
\end{tabular}
\caption{Parameter file commands used to run the code.}
\label{parameterTable}
\end{center}
\end{table}

\begin{figure}[h!]
\centering
\includegraphics[scale=0.22]{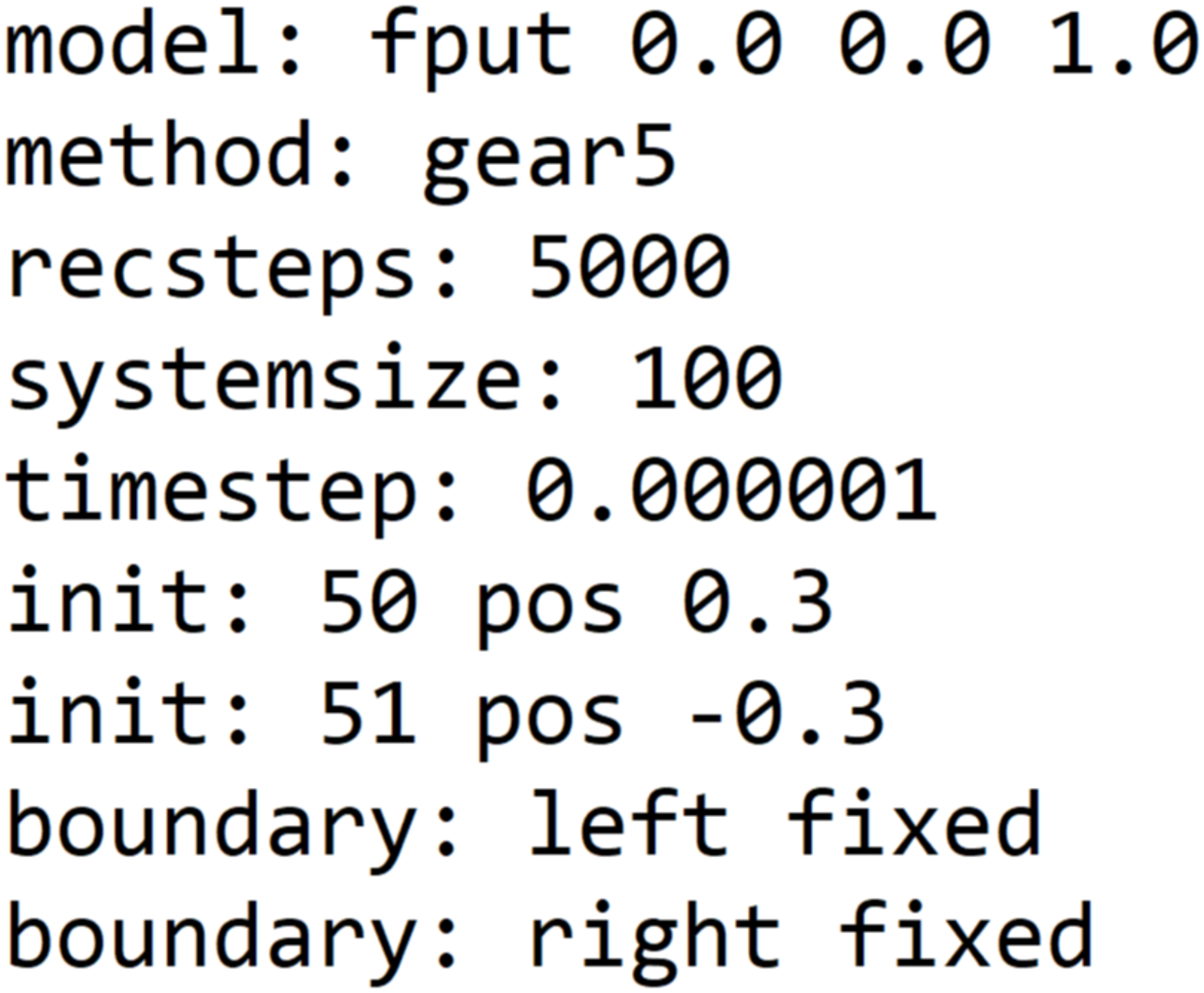}
\caption{An example of the parameter file used for the seeded localized nonlinear excitation in the FPUT chain is shown here. The results are discussed in Section \ref{fputResults}.}
\label{parameterLNE}
\end{figure}

One can set the type of potential and its parameters and choose from the integration algorithms provided. The user can also set various system and simulations parameters such as sampling intervals, mass of individual particles, chain sizes etc. as shown in Table. \ref{parameterTable}. An example of the parameter file is shown for the test case of the seeded localized nonlinear excitation in Section \ref{fputResults} is shown in Fig. \ref{parameterLNE}.

In addition to potential parameters and algorithm parameters, initial conditions for position, velocity and acceleration can be set on a particle by particle basis using either fixed numbers or random numbers within a range. They may also be imported from a file with the data stored in the format shown in Table. \ref{initTable}.

\begin{table}
\begin{center}
\begin{tabular}{ c c c }
\hline
 $x_1$ & $v_1$ & $a_1$  \\ 
 $x_2$ & $v_2$ & $a_2$  \\ 
 $x_3$ & $v_3$ & $a_3$  \\ 
 .     & .     & .      \\
 .     & .     & .      \\
 .     & .     & .      \\
 .     & .     & .      \\
 .     & .     & .      \\
 $x_N$ & $v_N$ & $a_N$  \\ 
\hline
\end{tabular}
\caption{Format for initial conditions to be read from file.}
\label{initTable}
\end{center}
\end{table}

The feature to import data from file may be also leveraged to continue runs from the last known point in time of a previous simulations recorded in the restart file. Finally, boundary conditions can be specified in the parameter file separately for the left and right boundaries. 

\section{Results: test cases}
\label{results}

\subsection{Toda lattice}
\label{todaResults}

In the limit of $N \rightarrow \infty$, the Toda lattice from Eq. \ref{todaPotential} admits a soliton solution of the following form \cite{toda1981}, where by soliton we mean a SW which does not interact with another SW except for a phase change suffered during the collision 
\begin{eqnarray}
x_n & = & \frac{1}{k_2}\ln\frac{1 + \exp(2(\kappa n - \kappa + \beta t))}{1 + \exp(2(\kappa n + \beta t))},
\label{todaSolitonX}
\end{eqnarray}
where $x_n$ refers to the displacement of the $n^{th}$ mass.

The velocity of the $n$th particle can be found by taking a derivative with time of $x_n$ and is 
\begin{eqnarray}
v_n & = & -\frac{2\beta \exp(2(\kappa n + 2\beta t))(\exp(2\kappa) - 1)}{k_2(1 + \exp(2(\kappa n + \beta t)))(\exp(2\kappa) +\exp(2(\kappa n + \beta t))}.
\label{todaSolitonV}
\end{eqnarray}

In the above solution, $\kappa$ is a free parameter which controls the width of the soliton, width $ \propto 1/\kappa$ and $\beta = \pm\sqrt{k_1k_2/m}\sinh{\kappa}$. Here, $\kappa > 1$. The soliton has the following features - as the soliton propagates it changes the global state of the system. Solitons of smaller widths propagate faster and the sign of $\beta$ controls the direction of soliton propagation. For our simulations, we used $k_1 = 1.0, k_2 = 10.0$ and varied $\kappa$ to seed soliton profiles into the Toda lattice.

We use Eq. \ref{todaSolitonX} and \ref{todaSolitonV} to seed the soliton profile into the Toda chain. The Toda lattice soliton solution holds true for an infinite system. Due to the fact that the soliton is kink shaped i.e., the right end is in a different configuration than the left end, the finite sized chain recoils as soon as the system is released. This leads to some radiative emissions from the soliton profile seeded into the chain (too small to be seen on the kinetic energy plot). In our simulations, $x_1 = 0$ and $x_N = -0.06$. Due to emission of this radiation, the soliton slows down slightly as compared to the theoretical solution. The effects of boundaries on the system dynamics is an important issue and has been the subject of previous study where the slowing down of the numerical solutions as compared to the theoretical solution has been reported \cite{shenPRE2014}. 

\begin{figure*}[tbp]
\centering
\includegraphics[scale=0.3]{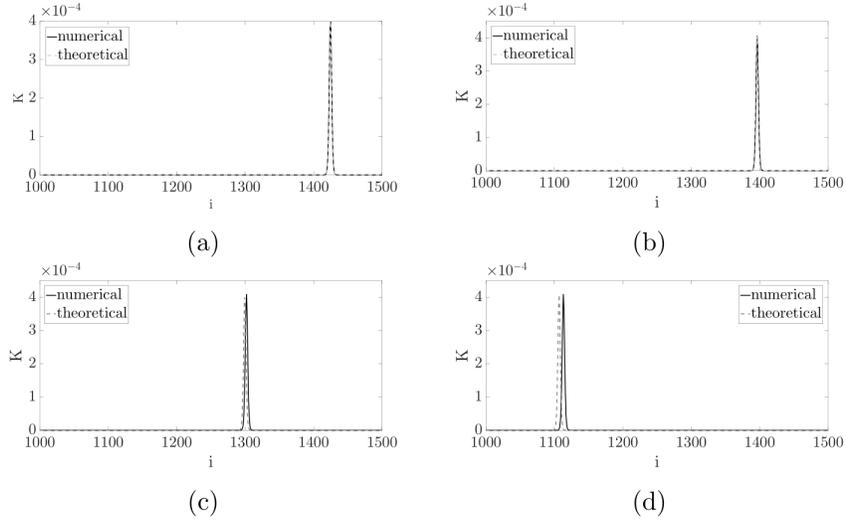}
  
    \caption{Subfigures (a) through (d) show the Toda soliton propagating on a lattice at times $t = 0, 10, 40$ and $100$ respectively. The numerical solution is shown as the solid line while the exact solution is shown as a dashed line.}
  \label{todaSwSingle}
\end{figure*}

In our simulations, we have used open boundaries and seeded the kink closer to the right boundary to minimize the recoil. The time-step used is 0.01 and the data is recorded at every 100 iterations, setting the time scale for the results shown in Fig. \ref{todaSwSingle}. The energy is conserved to one part in $10^6$ using a Gear 5th order predictor-corrector algorithm.

\begin{figure}
\centering
\includegraphics[scale=0.24]{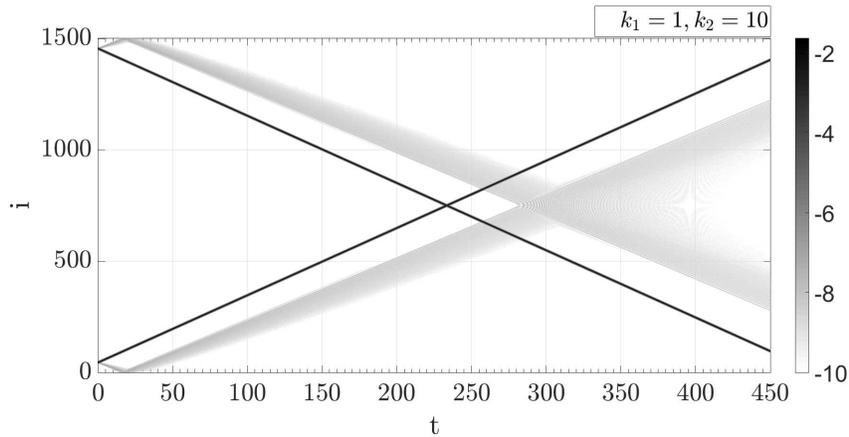}
\caption{The collision of two Toda solitons as they propagate on the lattice is shown in the panel. The figure shows the logarithm of kinetic energy plotted against time on the x-axis and particle index on the y-axis. The color sets the scale of the energy plotted with darker color corresponding to higher energy as shown in the color bar.}
\label{todaSwCollision}
\end{figure}

The Toda lattice also allows for multiple soliton solutions. We show in Fig. \ref{todaSwCollision} two solitons seeded in the chain at opposite ends and traveling towards each other. As soon as the simulation starts, the solitons radiate some energy as explained previously. However, after the radiation separates from the solitons, they move with constant speed towards each other and meet at the center of the chain. As expected, the two solitons show no scattering when they collide with each other. Further, the front of the radiation moves slightly slower than the solitons themselves. This is due to the fact that the soliton speed in the system is always greater than the speed of sound which is the fastest speed at which the radiation can propagate \cite{todaJPSP1967_1,todaJPSP1967_2,toda1981}. For the soliton collision problem we have used $\kappa_1 = \kappa_2 = 0.6, k_1 = 1.0$ and $k_2 = 10.0$. 

\subsection{FPUT lattice}
\label{fputResults}

The FPUT system is a nonlinear spring-mass model with a polynomial potential given by Eq. \ref{fputPotential}. The model was originally proposed and studied to show that in contrast to the finite sized harmonic oscillator chain, a nonlinear spring-mass chain would thermalize i.e., the energy of the modes would be shared and this would lead to the eventual equipartitioning of energy \cite{fermiLANLR1955,fermiChicago1965}. What they found however, was something they never expected. Their simulations showed that the energy stored in a single mode was shared with other modes for short times, but at later times, the energy would return back to the same mode i.e. the system showed no equipartition of energy.

Using PULSEDYN, we recreated the recurrence phenomenon observed by Fermi and coworkers. For our simulations, we used $k_1 = 0.5, k_2 = 0.0833$ and $k_3 = 0$ in a 64 particle system in Eq. 2 with $m_i = 1$. We initialized the system with all the energy fed into the lowest mode, $k = 1$ using the following equation, 
\begin{eqnarray}
x_i & = & \sum_{k = 1}^{N} x(k)\sin\frac{ik\pi}{N+1}.
\end{eqnarray}
The mode energies are then given by 
\begin{eqnarray}
E_k & = & \frac{1}{2}\dot{a}_k ^2 + \omega_k ^2 a_k ^2.
\end{eqnarray}
Here, 
\begin{eqnarray}
\omega_k ^2 & = & \frac{4k}{m}\sin^2{\frac{k\pi}{2(N+1)}},
\end{eqnarray}
and
\begin{eqnarray}
a_k & = & {\frac{2}{N}}\sum_{i = 1}^{N}x_i\sin{\frac{ik\pi}{N+1}}.
\end{eqnarray}

\begin{figure}
\centering
\includegraphics[scale=0.24]{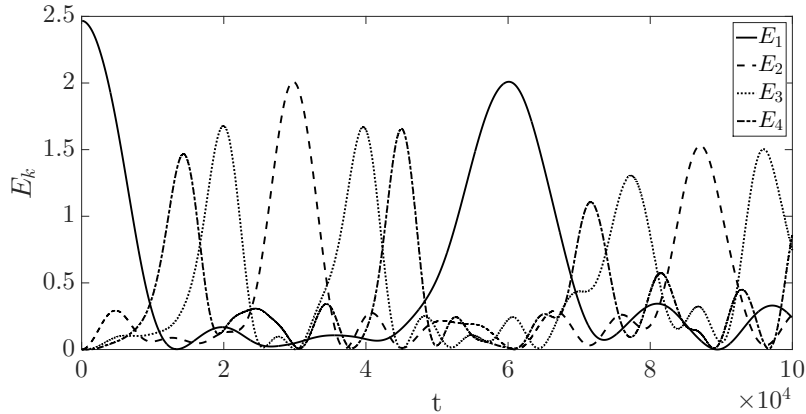}
\caption{The energy stored in each of the first 4 modes of the $\alpha$-FPUT chain as a function of time is shown here. The recurrence of energy in the first mode is seen at $t \approx 60000$. Here, $k_1 = 0.5, k_2 = 0.0833, k_3 = 0$ and the energy is initialized in the first mode of the system i.e. $k = 1$.}
\label{fputRecurrence}
\end{figure}

While the energy is initially shared with higher modes $k = 2, 3, ...$, it returns back to the original mode after some time as shown in Fig. \ref{fputRecurrence}. The total energy here is conserved to 1 part in $10^5$ using a time step of 0.1 with the velocity-Verlet algorithm \cite{swopeJCP1982}. We have used fixed boundaries for the system.

We now turn our attention to phenomena in the FPUT chains which have been studied more recently. It has been shown that a bond squeezing or stretching in the $\beta$ model of the FPUT chain i.e. $k_2 = 0$ leads to a localized nonlinear excitation (LNE) \cite{senPhysRev2009,kashyapIJMPB2017}. A squeeze in a bond is effected by setting the following initial condition, $x_{i} = -x_{i+1} = A$. The LNE is not stable and it delocalizes after some time and the time scale depends on the parameters of the system as well as the initial conditions \cite{kashyapIJMPB2017}. In the absence of phonons, i.e. $k_1 = 0$, it has been also shown that the destabilization takes place via emission of SWs and anti-SWs (ASWs) and other metastable nonlinear excitations \cite{kashyapIJMPB2017}. Following the destabilization of the LNE, the system then enters a state called quasi-equilibrium (QEQ) where the velocity distribution is Gaussian, but the kinetic energy fluctuations are too high for the state to be equilibrium \cite{senPhysRep2008,senPhysicaA2004,avalosPRE2014}. The potential in this case is given by

    \begin{eqnarray}
    V_{i, i+1} & = & k_3(x_{i+1} - x_{i})^4,
    \label{quarticFPUT}
    \end{eqnarray}

\begin{figure}
\centering
\includegraphics[scale=0.24]{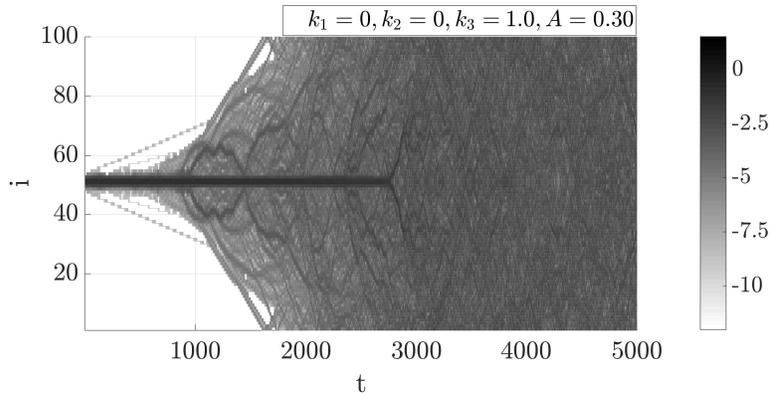}
\caption{A contour plot of $\log_{10}{KE}$ in the quartic FPUT system with $k_1 = 0$ is shown here. The LNE is in bond 51 of a 100 particle chain with fixed boundaries. Darker shades correspond to higher energy as shown in the color bar.}
\label{lneContourPlot}
\end{figure}

A simulation of a seeded LNE using the potential in Eq. \ref{quarticFPUT} is shown in Fig. \ref{lneContourPlot}. The results are as expected. The LNE decays by emitting SWs and other metastable excitations. A more detailed discussion of the LNE decay and the evolution of the system to equilibrium is presented in Section \ref{journeyToEqb}. The time step used here is $10^{-6}$. A small time step is required for a problem such as this one owing to the high frequency quasi-periodic oscillations of the LNE before delocalization. The energy is conserved to better than one part in $10^{11}$ using the velocity-Verlet algorithm \cite{kashyapIJMPB2017}. 

\begin{figure}[h!]
\centering
\includegraphics[scale=0.5]{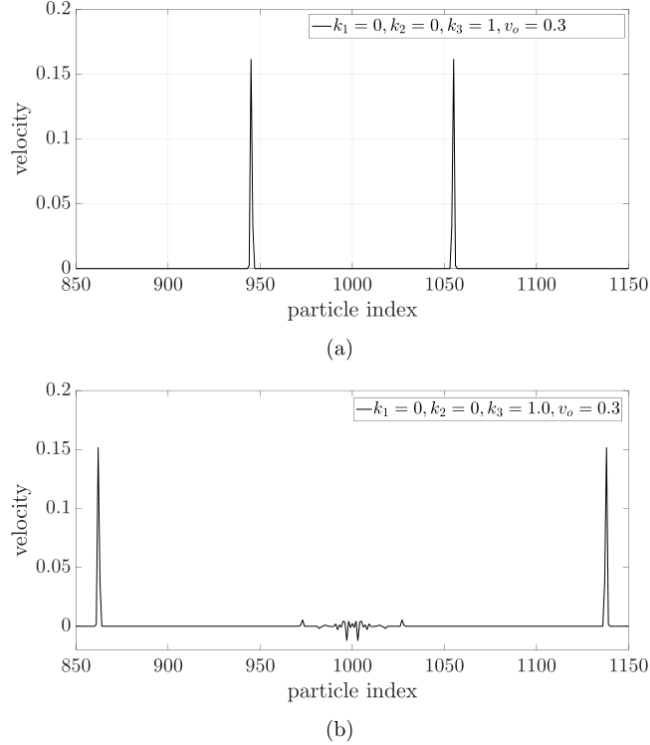}
\caption{Subfigure (a) shows two SWs in the quartic FPUT system traveling towards each other at $t = 1150$. Subfigure (b) shows the two SWs moving away from each other towards the ends of the chain at $t = 1560$ after colliding with each other. The collision takes place at particle 1000.}
\label{fputSwCollision}
\end{figure}

Another test for the quartic FPUT system is the dynamics of SW collisions in the system. A SW can be seeded in the chain by giving a velocity perturbation to a particle at $t = 0$ and letting it evolve. We seed two SWs in a quartic system (with $k_1=k_2=0$). The SWs are identical and propagate towards each other from opposite sides of the chain. It has been shown that two such interacting waves experience scattering when they meet and they leave behind residual energy \cite{zhaoPRL2005}. The results are shown below for a 2000 particle chain in Fig. \ref{fputSwCollision}. Two SWs were created at particle 400 and 1600 by giving them velocity perturbations of strength $v_o = 0.3$. The boundaries are set to be free boundaries and the simulation was performed with the velocity-Verlet algorithm with a time step of $10^{-3}$. The total energy of the system is conserved to 1 part in $10^7$. At particle 1000 where the two SWs meet they collide and leave behind oscillations as expected. Due to energy conservation, the heights of the SWs post collision are slightly reduced. If this process happens repeatedly, the SWs break up into many secondary SWs and the system enters the QEQ phase which is similar to the equilibrium phase except that energy equipartitioning is not seen \cite{senPhysRep2008}. 

\subsection{Morse and Lennard-Jones potentials}
\label{morseLjResults}

Exact solutions of the Morse and the Lennard-Jones (LJ) chains in 1D do not exist to our knowledge. Therefore, to test propagation of solitons through chains with the Morse and LJ potential, we follow the numerical scheme employed by Flytzanis \textit{et al} \cite{flytzanisJPA1989}. In their 1989 study, the authors reduce the equations of each of the potentials under consideration into their corresponding $\alpha+\beta$ FPUT forms as polynomial expansions of these nonlinear potentials. The corresponding Boussinesq equation of the FPUT chain is derived. The Boussinesq equation has well known soliton solutions. These solutions are used as soliton \textit{generators} to seed solitons into the chains. The procedure is as follows.

Consider the Morse potential for instance given by Eq. \ref{morsePotential}.
    
This potential can be expanded and written as a truncated polynomial potential of the form

\begin{eqnarray}
V_{i+1,i} & = & \frac{1}{2}G(x_{i+1} - x_i)^2 + \frac{1}{3}A(x_{i+1} - x_i)^3 + \frac{1}{4}B(x_{i+1} - x_i)^4.
\label{fputExp}
\end{eqnarray}

In the continuum limit, the right hand side of Eq. \ref{fputExp} can be expanded as 

\begin{eqnarray}
V'(r_{i+1}) - 2V'(r_i) + V'(r_{i-1}) & \approx & D^2 \frac{\partial^2}V{\partial y^2} + \frac{1}{12}D^4\frac{\partial^4V}{\partial y^4}.
\end{eqnarray}

Here, $r_i = x_{i+1} - x_i$ and $y = iD$ where $D$ is the bond length of each bond and $i$ is the site index. The corresponding Boussinesq equation takes the form

\begin{eqnarray}
u_{tt} - c_o^2u_{yy} - p(u^2)_{yy} - q(u^3)_{yy} - hu_{yyyy} & = & 0,
\end{eqnarray}
with $c_o^2 = GD^2/m, p = AD^3/m, h = GD^4/(12m)$ and $q = BD^4/m$. The soliton solution of this equation is given by 

\begin{eqnarray}
y(x, t) = \pm 2 sgn(h)\Big (\frac{2h}{q}\Big )^{(1/2)}\arctan{\Big ( \frac{1}{w}\tanh\Big(\frac{x-vt}{L} \Big ) \Big )},
\label{solitonGen}
\end{eqnarray}
where
\begin{eqnarray}
w & = & \Bigg (\frac{[4p^2 + 18(v^2 - c_o^2)q]^{1/2} \pm 2p}{[4p^2 + 18(v^2 - c_o^2)q]^{1/2} \mp 2p}\Bigg ),
\end{eqnarray}
and 
\begin{eqnarray}
L = 2\Big [\frac{h}{v^2 - c_o^2}\Big]^{1/2}.
\end{eqnarray}

The soliton solution of the Boussinesq equation is then used as a soliton generator to seed soliton shapes into the discrete chains. Since this solution is not a true solution of the discrete system, the solution decomposes into a radiative part and a soliton-like kink part. In the study \cite{flytzanisJPA1989}, the authors found that the propagating kink soliton does not fit well to polynomial fits and other simple fits. In fact, the stable part of the propagating kink is fit best to a generalized Toda soliton given by the equation
\begin{eqnarray}
x_n & = & \frac{1}{2}A^w\log{\frac{1 + \exp[\pm 2(n - n_o -0.5)/w]}{1 + \exp[{\pm 2(n - n_o + 0.5)/w]}}} + C,
\label{todaSolnGen}
\end{eqnarray}
where, $A$ is the height of the kink, $n_o$ is the position of the center of the kink and $2w$ is the width of the soliton kink. Flytzanis \textit{et al} \cite{flytzanisJPA1989} were able to obtain very good fits of the non-radiative part of the propagating kink in the LJ and the Morse chains using the above forms for chains with $N >> 1$.

In this section, we use the same approach to seed solitons in the Morse and LJ potentials and verify that the propagating constant velocity front of the kink does indeed behave as a Toda soliton. 

\begin{figure}[h!]
\centering
\includegraphics[scale=0.5]{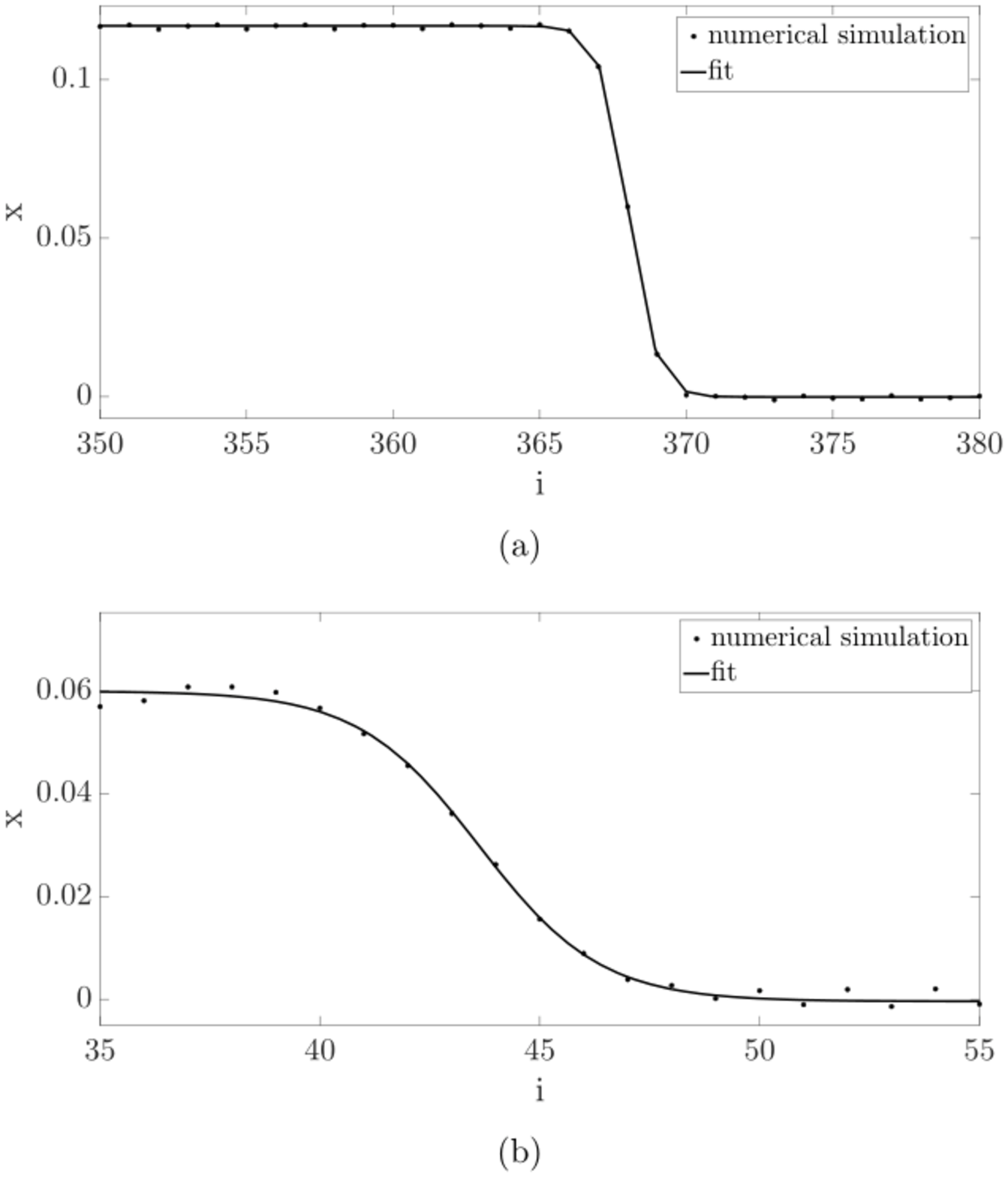}
\caption{Subfigure (a) shows a snapshot of the kink profile in Morse chain at $t = 310$ and the fit to the displacements from Eq. \ref{todaSolnGen}. Subfigure (b) shows a snapshot of the kink at $t = 54$ of the LJ chain. The data points are obtained from numerical simulations and the line shows the fit to Eq. \ref{todaSolnGen}.}
\label{morseLjTodaFit}
\end{figure}

For the Morse potential, we used the parameters $k_1 = 0.01$ and $k_2 = 7$. This leads to the values of $G = 1, A = -10.5$ and $B = 57$. For the LJ potential, we use $k_2 = 0.01389$ and $k_1 = 1$ to obtain values of $G = 1, A = -10.5$ and $B = 62$. We use these values of $G, A$ and $B$ to calculate $p, q, h$ and $c_o^2$ after setting $D = 1$. Then we use the values of $p, q, h$ and $c_o^2$ to calculate the soliton generator from Eq. \ref{solitonGen} and seed this shape into the Morse and the LJ chains respectively. We track the propagating kink and obtain the displacement of each particle in the chain at a later time and fit it to the generalized Toda lattice solution from Eq. \ref{todaSolnGen}. The time step used is 0.1 and the energy conservation holds to 1 part in $10^5$. The plots are shown in Fig. \ref{morseLjTodaFit}.

As with Flytzanis' study, the fit obtained is satisfactory. For the Morse potential, we get the parameters of the fit as $A = 0.117 \pm 0.005, w = 1.00 \pm 0.02$ and $n_o = 368$. The goodness of the fit $R^2 = 0.999$. For the case of the LJ chain we found, $A = 0.22 \pm 0.04, w = 2.7 \pm 0.6, n_o = 43.6 \pm 0.4$ and $R^2 = 0.985$. 

\section{Results: Journey to Equilibrium}
\label{journeyToEqb}

\begin{figure}
\centering
\includegraphics[scale=0.5]{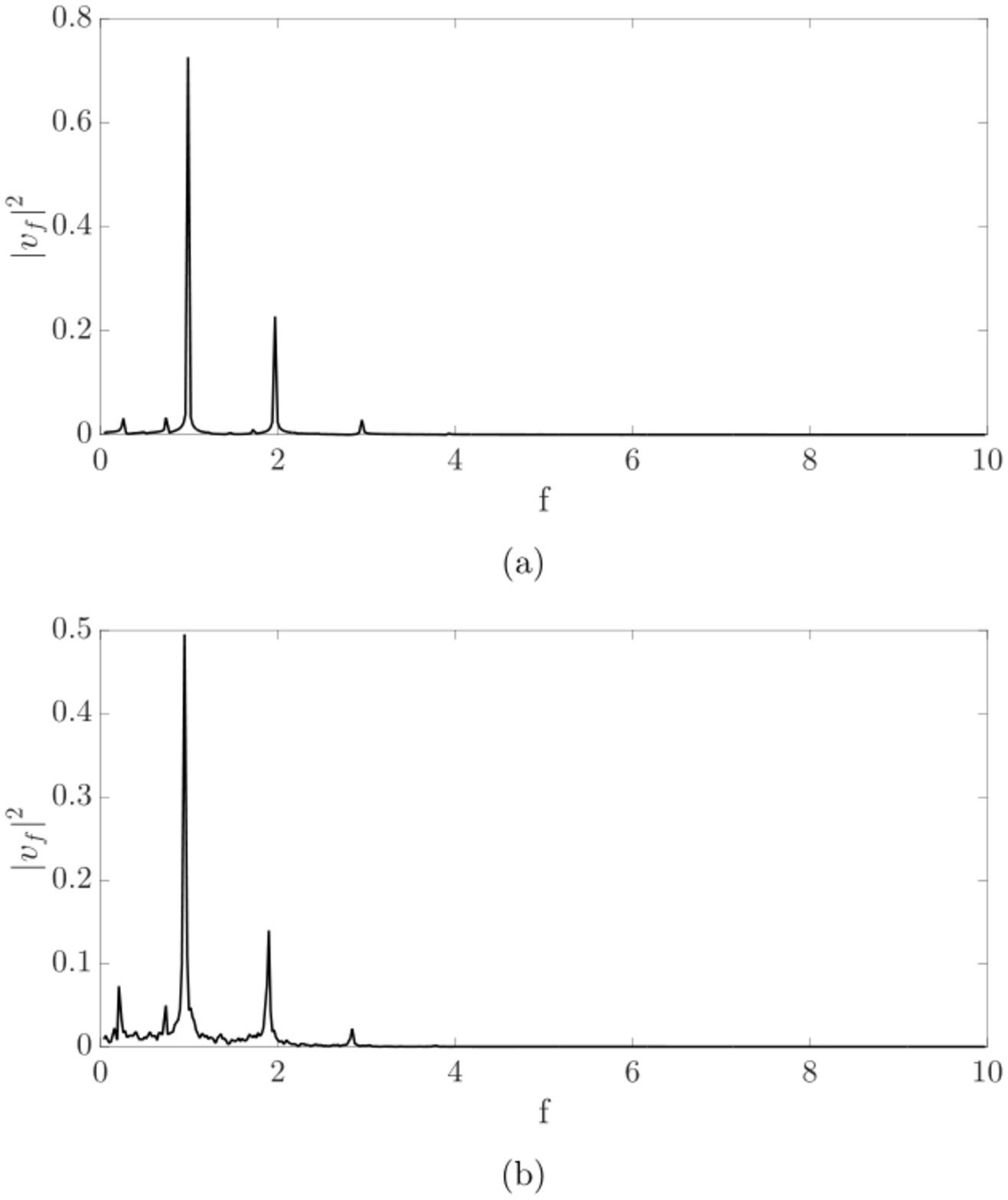}
\caption{Shown here are the direct cosine transforms (DCTs) of the kinetic energy of the LNE particle (particle 50) at two different time intervals in its evolution. Subfigure (a) shows the DCT for the time interval $t = 0$ to $t = 100$ and Subfigure (b) shows the DCT computed for the time interval from $t = 2500$ to $2600$ for the parameter values $k_1 = 0, k_2 = 0, k_3 = 1.0$ and $A = 0.30$. At early times, the DCTs show sharper peaks. With time, as the LNE delocalizes, the DCT shows frequency spreading i.e. the LNE is not quite periodic anymore.}
\label{lneDctSpread}
\end{figure}

As demonstrated in the previous section, a seeded LNE eventually delocalizes. As it delocalizes, it does so in a way that system energy distribution goes towards the value predicted by the virial theorem. At early times in the LNE evolution, the motion is very nearly periodic with the frequencies corresponding to those of the Duffing oscillator \cite{kashyapIJMPB2017}. As it evolves and emits energy via SWs and other excitations, the frequency peaks broaden as shown in Fig. \ref{lneDctSpread}, i.e., the LNE's periodicities are not as well defined. This accelerates the delocalization process. Once delocalization is complete the system enters a QEQ state which turns out to be a metastable state \cite{senPhysRev2009,avalosPRE2009,avalosPRE2014}. 

\begin{figure}[h!]
\centering
\includegraphics[scale=0.5]{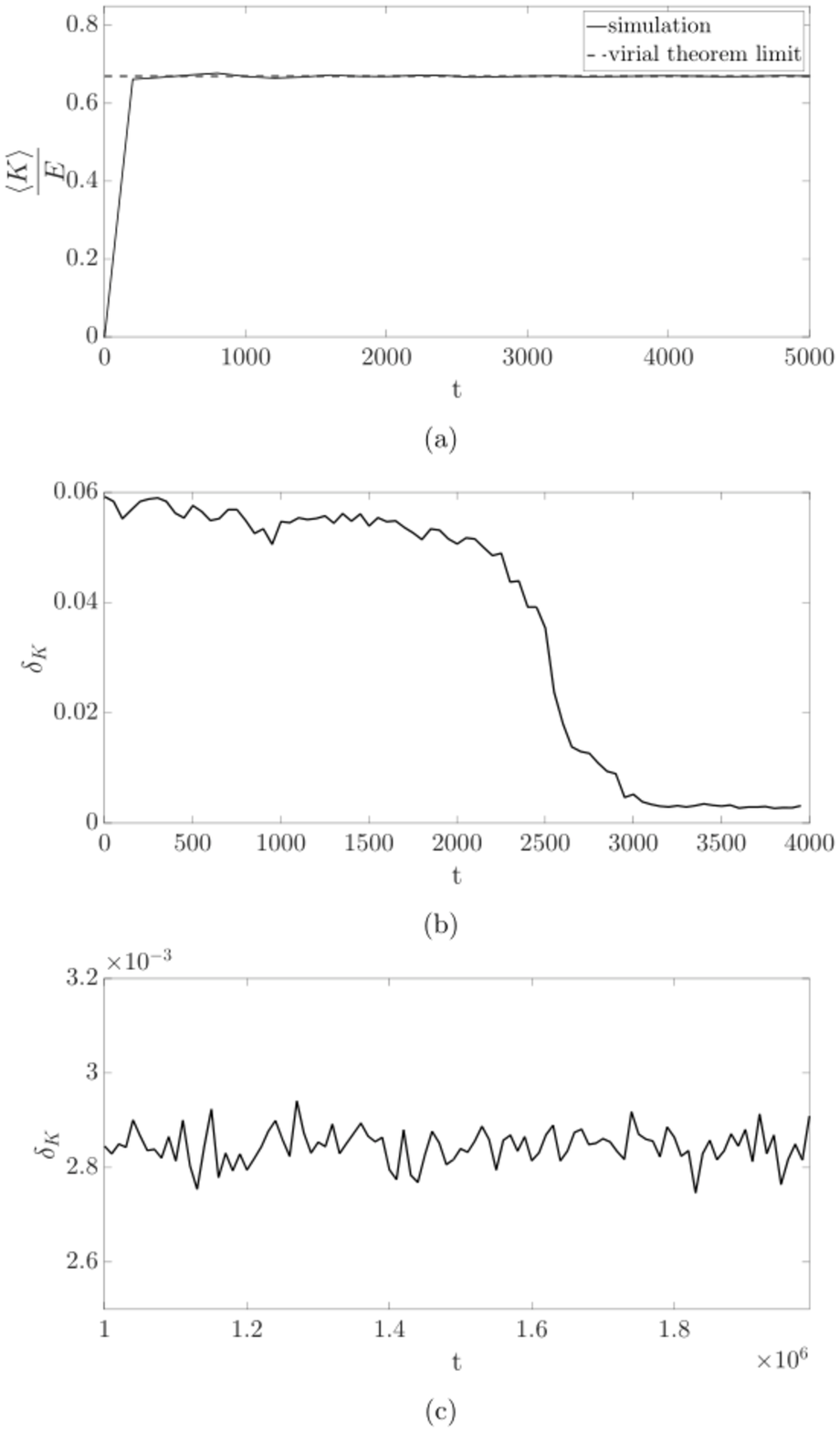} 
\caption{Subfigure (a) shows the ratio of the average kinetic energy to the total energy of the system. LNE delocalization is complete here at $t = 4000$. Subfigure (b) shows the fluctuations in $\langle K\rangle$ as a function of time from $t = 0$ to $t = 4000$. By $t = 4000$, the LNE delocalizes and there is a corresponding decrease in the fluctuations as the system moves from a quasi-periodic behavior into the QEQ phase. Subfigure (c) shows the fluctuations at a much later time when the system has reached equilibrium. Here, the parameters are $k_1 = 0, k_2 = 0, k_3 = 1.0$ and the initial conditions are a seeded LNE at particle 50 and 51 with $x_{50} = -x_{51} = A = 0.3$.}
\label{lneFluctuations}
\end{figure}

As the LNE delocalizes, the kinetic energy fluctuations $\delta_K$ of the system begin to decrease since the motion of the LNE particles is not nearly as periodic as at early times. A plot of $\delta_K$ fluctuations during the LNE lifetime is shown in Fig. \ref{lneFluctuations}(b). With time, the fluctuations go down further and reach a stable value as shown in Fig.\ref{lneFluctuations}(c). This stable regime is well after the delocalization of the LNE. The question now arises - is the late time state true equilibrium or is it QEQ? In true equilibrium, ergodicity would hold true and the energy would be equipartitioned. We know that the answer to this question is known, namely the FPUT chain goes to the equipartitioned state at late times for weak nonlinearity \cite{liviPRA1985,onoratoPNAS2015,benettinJSP2013}. Our objective hence is to make sure that PULSEDYN can deliver the anticipated conclusion for a system in the nonlinear case (recall $k_1=k_2=0$).

To address this question, we performed calculations to verify the virial theorem and the specific heat of the system. The parameters used are $k_1 = 0, k_2 = 0, k_3 = 1.0$ and $A = 0.3$. These are the same parameters as those for the LNE in Fig. \ref{lneContourPlot}. We use a time step of $10^{-5}$ after LNE delocalization and run the simulation for $10^{12}$ time iterations. We sample data at every $10^5$ iterations and we have called each such sampling a single time step. We have collected we have recorded data in this case for $1.4\times 10^7$ time steps. 

We find that at late times - the last $15\%$ of the time evolution - the space average and the time average of $K$ are equal to each other, i.e., the system is ergodic. We then check whether the virial theorem results are satisfied. The virial theorem states 
\begin{eqnarray}
\langle K\rangle & = & \frac{n}{n+2}E.
\end{eqnarray}
Here, $E$ is the total energy of the system and $n$ is the exponent in the power law potential. For the purely nonlinear $\beta$-FPUT chain, $n = 4$. Therefore, $\langle K\rangle = \frac{2}{3}E$, which we recover from our calculations shown in Fig. \ref{lneFluctuations}(a). However, ergodicity and the virial theorem are both satisfied even in the QEQ state. Further, the velocity distribution in the QEQ state has been found to be Gaussian \cite{kmPramana2005}. Therefore, to establish conclusively that the final state obtained is truly equilibrium, the system has to satisfy equipartition. 

Calculations of response functions in the microcanonical ensemble in a nonlinear many body system are challenging \cite{scalasPRE2015,rayPRA1991}. To examine whether the FPUT system has reached the equipartitioned state, we rely on a recent similar study carried out for an alignment of grains where the grains repel via a strongly nonlinear algebraic potential \cite{michellePRE2017}. There it has been shown that 
\begin{eqnarray}
C_v & \approx & k_B\Bigg[\frac{n+2}{2n} - \frac{1}{N}\Bigg(\frac{n+2}{n} + \frac{4(N-2)}{2nN}\Bigg)\Bigg].
\label{cvFput}
\end{eqnarray}
Here, $N$ is the number of particles in the finite sized system. In the above equation, if $N \rightarrow \infty$, we recover the result from Tolman's generalized equipartition theorem in Eq. \ref{tolmanCv} due to the equivalence of statistical ensembles in the thermodynamic limit \cite{tolmanPR1918} as 
\begin{eqnarray}
C_v & \approx & \Bigg (\frac{n+2}{2n}\Bigg )k_B.
\label{tolmanCv}
\end{eqnarray}
Since the result depends only on the exponent in the potential $n$, Eq. \ref{cvFput} holds for the purely nonlinear $\beta$-FPUT chain when the energy is in the equipartitioned state. 

For $N = 100$, the theoretical value for $C_v/k_B$ from Eq. \ref{cvFput} is calculated to be 0.725. To compare this value to $C_v$ that can be extracted from our simulations we need to find a relation between the kinetic energy fluctuations and $C_v$. Such a connection has recently been established in \cite{rughJPAMG1998},
\begin{eqnarray}
C_v & = & \frac{k_B}{N}\Bigg(1 - \frac{(N-4)\langle 1/K^2 \rangle}{(N-2)\langle 1/K^2 \rangle}\Bigg)^{-1}.
\label{cvNum}
\end{eqnarray}
From our numerical calculations, we recover a value of 0.729 at late times using Eq. \ref{cvNum}. The numerical value from our simulations is within $0.55\%$ of the theoretical value and shows agreement between the theoretical estimate and the dynamical simulation based results. Therefore, our studies suggest that the purely nonlinear $\beta$-FPUT system goes to the equipartitioned state at late times in this study, which is a prediction based on the results of the PULSEDYN code. 

\section{Conclusions}
\label{conclusions}

We have presented the PULSEDYN code that is designed to make dynamical simulations of energy conserved and driven dissipative $1D$ nonlinear systems accessible to students and professionals without any background in the field. The code provides features that can be readily tuned. PULSEDYN is capable of carrying out detailed particle dynamics simulations in windows and linux platforms for an integrable system, the Toda chain, and three non-integrable systems, the FPUT, Morse and Lennard-Jones chains. The conservative systems can be studied using the velocity Verlet algorithm and the Gear algorithm whereas the driven dissipative system can only be studied using the Gear algorithm. The program has been written with a modular design and the writing style is expected to allow for additions and extensions to the code for an experienced $C++$ programmer.  

To explore the capabilities of PULSEDYN we have recovered the following results. We have recovered the exact analytical results on the propagation of a soliton and the interaction between solitons in the Toda lattice, we have confirmed the presence of the recurrence phenomenon, we have explored the problem of LNE delocalization and SW interactions in the FPUT chain and our results are in agreement with known results in the FPUT chain regarding SW interactions, and we have recovered the approximate forms of SWs in the Morse and the LJ potential that have been previously investigated. 

In addition, we have also presented new results regarding the relaxation processes of LNEs in the purely nonlinear FPUT chain. We demonstrated that as the LNE delocalizes the fluctuations in the system's kinetic energy $K$ goes down at late enough times as expected. At sufficiently late times the energy distribution reaches the values predicted by the virial theorem and the system is found to reach the equipartitioned state. The equipartitioning is evident through the excellent agreement between the specific heat calculated from the simulations and the same predicted theoretically.

\section*{Acknowledgements}
\label{acknowledgements}
We thank Tyler Barrett, Guo Wen, Kevin VanSlyke and Alexandra Westley for their help in testing the PULSEDYN code for various cases. Rohith Kashyap's advice during the development of the code is gratefully acknowledged.

\clearpage
\section*{Metadata}
\label{}

\section*{Current executable software version}
\label{}

\begin{table}[!h]
\begin{tabular}{|l|p{6.5cm}|p{6.8cm}|}
\hline
\textbf{Nr.} & \textbf{(executable) Software metadata description} & \textbf{Please fill in this column} \\
\hline
S1 & Current software version & v1.0\\
\hline
S2 & Permanent link to executables of this version  & https://github.com/rahulkashyap7557\\
\hline
S3 & Legal Software License & GNU General Public License 3 (GPL) \\
\hline
S4 & Computing platform/Operating System & Windows, Linux \\
\hline
S5 & Installation requirements \ & none \\
\hline
S6 & If available, link to user manual - if formally published include a reference to the publication in the reference list & https://github.com/rahulkashyap7557\\
\hline
S7 & Support email for questions & rahulkashyap7557@gmail.com\\
\hline
\end{tabular}
\caption{Software metadata}
\label{} 
\end{table}

\pagebreak

\section*{Current code version}
\label{}
\begin{table}[!h]
\begin{tabular}{|l|p{6.5cm}|p{6.8cm}|}
\hline
\textbf{Nr.} & \textbf{Code metadata description} & \textbf{Please fill in this column} \\
\hline
C1 & Current code version & v1.0 \\
\hline
C2 & Permanent link to code/repository used of this code version & https://github.com/rahulkashyap7557\\
\hline
C3 & Legal Code License   & GNU General Public License 3 (GPL) \\
\hline
C4 & Code versioning system used & github \\
\hline
C5 & Software code languages, tools, and services used & C++ \\
\hline
C6 & Compilation requirements, operating environments \& dependencies & Code::Blocks, GNU GCC compiler \\
\hline
C7 & If available Link to developer documentation/manual &https://github.com/rahulkashyap7557\\
\hline
C8 & Support email for questions & rahulkashyap7557@gmail.com\\
\hline
\end{tabular}
\caption{Code metadata}
\label{} 
\end{table}


\begin{thebibliography}{00}

\bibitem{fermiLANLR1955} E. Fermi, J. Pasta, and S. Ulam, Los Alamos Report LA-1940, 1955.
\bibitem{fermiChicago1965} E. Fermi, \textit{Collected Papers}, vol. II (Univ. of Chicago Press, 1965).
\bibitem{todaJPSP1967_1} M. Toda, J. Phys. Soc. Jpn. 22 (1967) 431-436.
\bibitem{todaJPSP1967_2} M. Toda, J. Phys. Soc. Jpn. 23 (1967) 501-506.
\bibitem{flachPhysRep1998} S. Flach and C. R. Willis, Phys. Rep. 295 (1998) 181-264.
\bibitem{flachPhysRep2008} S. Flach and A. V. Gorbach, Phys. Rep. 467 (2008) 1-116.
\bibitem{bermanChaos2005} G. P. Berman and F. M. Izrailev, Chaos 15 (2005) 015104.
\bibitem{campbellChaos2005} D. K. Campbell, P. Rosenau and G. M. Zaslavksy, Chaos 15 (2005) 015101.
\bibitem{fordPhysRep1992} J. Ford, Phys. Rep. 213 (1992) 271-310.
\bibitem{saitoJPSP1967} N. Saito and H. Hirooka, J. Phys. Soc. Jpn. 23 (1967) 167-171.
\bibitem{reigadaPRE2001} R. Reigada, A. Sarmiento, and K. Lindenberg, Phys. Rev. E 64 (2001) 066608.
\bibitem{reigadaPhysicaA2002} R. Reigada, A. Sarmiento, and K. Lindenberg, Physica A 305 (2002) 467-485.
\bibitem{sieversPRL1988} A. J. Sievers and S. Takeno, Phys. Rev. Lett. 61 (1988) 970-973.
\bibitem{satoNature2004} M. Sato and A. J. Sievers, Nature 432 (2004) 486-488.
\bibitem{bennetinJSP2011} G. Bennetin and A. Ponno, J. Stat. Phys. 144 (2011) 793-812.
\bibitem{takatoEPL2012} Y. Takato and S. Sen, Europhys Lett.100 (2012) 24003.
\bibitem{avalosPRE2011} Edgar Avalos, Diankang Sun, Robert L. Doney, and Surajit Sen, Phys. Rev. E 84 (2011) 046610.
\bibitem{senPhysRev2009} S. Sen and T. R. Krishna Mohan, Phys. Rev. E 79 (2009) 036603.
\bibitem{senPhysRep2008} S. Sen, J. Hong, J. Bang, E. Avalos and K. L. Doney, Phys. Rep. 462 (2008) 21-63.
\bibitem{flytzanisJPA1989} N. Flytzanis, St. Pnevmatikos and M. Peyrard, J. Phys. A: Math. Gen. 22 (1989) 783-801. 
\bibitem{gear1966} C.W. Gear, The numerical integration of ordinary differential equations of various orders, Technical Report 7126, Argonne National Laboratory (1966).
\bibitem{swopeJCP1982} W. C. Swope, H. C. Anderson, P. H. Berens and K. R. Wilson, J. Chem. Phys. 76 (1982) 637.
\bibitem{verletPR1967} L. Verlet, Phys. Rev. 159 (1967) 98-103.
\bibitem{stormer1921}C. St\"ormer, M\'ethode d'int\'egration num\'erique des \'equations diff\'erentielles ordinaires , C.R. Congress Internat. Strassbourg 1920 (1921) 243.
\bibitem{allen1987} M.P. Allen and D.J. Tildesley, Computer Simulation of Liquids, Clarendon, Oxford, 1987.
\bibitem{toda1981} M. Toda, Theory of nonlinear Lattices, Springer-Verlag Berlin, Heidelberg, Berlin, (1981).
\bibitem{kashyapIJMPB2017} R. Kashyap, A. Westley, A. Datta and S. Sen, Int. J. Mod. Phys. B 31 (2017) 1742014.
\bibitem{zhaoPRL2005} H. Zhao, Z. Wen,Y. Zhang and D. Zheng, Phys. Rev. Lett. 94 (2005) 025507.
\bibitem{avalosPRE2009} E. Avalos and S. Sen, Phys. Rev. E 79 (2009) 046607.
\bibitem{kmPramana2005} T. R. K. Mohan and S. Sen, Pramana 64 (2005) 423-431.
\bibitem{liviPRA1985} R. Livi, M. Petini, S. Ruffo, M. Sparpaglione and A. Vulpiani, Phys. Rev. A 31 (1985) 1039.
\bibitem{onoratoPNAS2015} M. Onorato, L. Vozella, D. Proment and Y. V. Lvov, Proc. Natl. Acad. Sci. 112 (2015) 4208.
\bibitem{benettinJSP2013} G. Benettin, H. Christodoulidi and A. Ponno, J. Stat. Phys. 152 (2013) 195.
\bibitem{fergusonJCP1982} W. E. Ferguson, H. Flaschka and D. W. McLaughlin, J. Comput. Phys. 45 (1982) 157.
\bibitem{henonPRB1974} M. H\'enon, Phys. Rev. B 9 (1974) 1921.
\bibitem{press1988} W. H. Press, S. A. Teukolsky, W. T. Vetterling and B. P. Flannery, Numerical Recipes: The Art of Scientific Computing, Cambridge University Press, New York, 1985.
\bibitem{morsePR1929} P. M. Morse, Phys. Rev. 34 (1929) 57-64.
\bibitem{lennardjonesPRSL1924} J. E. Lennard-Jones, Proc. R. Soc. Lond. A 106 (1924) 463-477. 

\bibitem{shenPRE2014} Y. Shen, P. G. Kevrekidis, S. Sen and A. Hoffman, Phys. Rev. E 90 (2014) 022905.
\bibitem{senPhysicaA2004} S. Sen, T. K. Mohan and J. M. Pfannes, Physica A 342 (2004) 336-343.
\bibitem{avalosPRE2014} E. \'Avalos and S. Sen, Phys. Rev. E 89 (2014) 053202.
\bibitem{scalasPRE2015} E. Scalas, A. T. Gabriel, E. Martin and G. Germano, Phys. Rev. E 92 (2015) 022140.
\bibitem{rayPRA1991} J. R. Ray and H. W. Graben, Phys. Rev. A 44 (1991) 6905-6908.

\bibitem{michellePRE2017} M. Przedborski, S. Sen and T. A. Harroun, Phys. Rev. E 95 (2017) 032903.
\bibitem{tolmanPR1918} R. C. Tolman, Phys. Rev. 11 (1918) 261-275.
\bibitem{rughJPAMG1998} H. H. Rugh, J. Phys. A Math. Gen. 31 (1998) 7761-7770.


\end{thebibliography}
\end{document}